\documentclass[twocolumn,showpacs,amsmath,amssymb]{revtex4}
\usepackage{appendix}
\usepackage{graphicx}
\usepackage{dcolumn}
\usepackage{bm}
\def\b{\begin{equation}}
\def\e{\end{equation}}
\begin{document}
\title{Strongly direction-dependent magnetoplasmons in mixed Faraday-Voigt configurations}

\author{Afshin Moradi$^{1*}$ }\author{Martijn Wubs$^{2,3,4}$}
 \altaffiliation[]{}
 \email{a.moradi@kut.ac.ir; mwubs@fotonik.dtu.dk} 
\affiliation{%
\emph{$^{1}$Department of Engineering Physics, Kermanshah University of Technology, Kermanshah, Iran\\ $^{2}$Department of Photonics Engineering, Technical University of Denmark, DK-2800 Kgs. Lyngby, Denmark \\ $^{3}$Center for Nanostructured Graphene, Technical University of Denmark, DK-2800 Kgs. Lyngby, Denmark \\ $^{4}$NanoPhoton - Center for Nanophotonics, Technical University of Denmark, DK-2800 Kgs. Lyngby, Denmark }
}%

\begin{abstract}

The electrostatic theory of magnetoplasmons on a semi-infinite magnetized electron gas is generalized to mixed Faraday-Voigt configurations. We analyze a new type of electrostatic surface waves that is strongly direction-dependent, and may be realized on narrow-gap semiconductors in the THz regime. A general expression for the dispersion relation is presented, with its dependence on the magnitude and orientation of the applied magnetic field. Remarkably, the  group velocity is always perpendicular to the phase velocity. Both velocity and energy relations of the found magnetoplasmons are discussed in detail. In the appropriate limits the known magnetoplasmons in the higher-symmetry Faraday and Voigt configurations are recovered.
\end{abstract}

 \maketitle

\section{Introduction}
It is well known that a static magnetic field causes various important changes in the electromagnetic behavior of different media~\cite{S.A.H.G1158, Y.Y554, S.B674}. Also, since the early works by Chiu and Quinn~\cite{K.W.C4707, K.W.C1}, it is known that on the surface of a semi-infinite magnetized cold electron gas, a surface magnetoplasmon (SMP) can oscillate at constant frequencies only. 
This infinite flat-band dispersion relation holds in the electrostatic approximation~\cite{Maier:2007a}, and as long as spatial dispersion (a.k.a. \textquotedblleft nonlocal response'') can be neglected~\cite{Raza:2015a,Fitzgerald:2016a}. 

\begin{figure}[tb]
  \centering
 \includegraphics[width=7cm,clip]{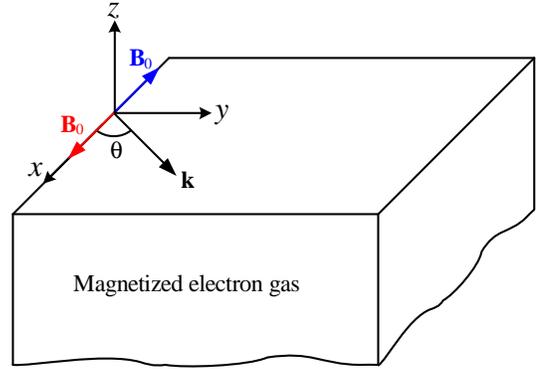}
  \caption{Sketch of a semi-infinite magnetized electron gas with static magnetic field $\textbf{B}_{0}$ parallel to the surface. Here we choose the direction of $\textbf{B}_{0}$ fixed along the $+\textbf{e}_{x}$ (red vector) or $-\textbf{e}_{x}$ (blue vector). Note that $\textbf{e}_{x}$ is the unit vector along the $x$-axis. Also, $\textbf{k}$ is the wavevector of the propagating surface wave, and $\theta$ (or $180^{\circ}-\theta$) is the angle between $\textbf{k}$ and $\textbf{B}_{0}$. The special cases  $\theta=0$ or $180^{\circ}$ and $\theta=90^{\circ}$ are called the Faraday and Voigt configurations, respectively. Here we study magnetoplasmons for the general case of arbitrary $\theta$, which we call mixed Faraday-Voigt configurations. Note that in the Voigt configuration the magnetoplasmons travel in the $+\textbf{e}_{y}$ direction. The electron gas is bounded from above by a semi-infinite insulator at $z=0$.   }
  \label{fig.1}
\end{figure}

In particular, there are two configurations for which this constant SMP frequency is given by 
$\omega=\sqrt{\omega_{\mathrm{p}}^{2}+\omega_{\mathrm{c}}^{2}}/\sqrt{2}$~\cite{H.J.L755}, where $\omega_{\mathrm{p}}$ is the electron plasma frequency and $\omega_{\mathrm{c}}= e B/m_e$ is the electron cyclotron frequency ($e$ and $m_{e}$ are the elementary charge and the electron mass, respectively).
The first of these configurations is the Faraday configuration, when the applied  magnetic field is parallel both to the surface and to
the direction of propagation of the wave (see sketch in Fig.~\ref{fig.1}). 
The other is the so-called  perpendicular configuration, when the applied out-of-plane magnetic field points perpendicularly both to the surface and to the propagation direction of the wave (not shown in Fig.~\ref{fig.1}). 
Interestingly, when the SMP in the Faraday configuration has strong surface-wave characteristics, then the SMP in the perpendicular configuration acts like a bulk wave, and vice versa~\cite{A.M}. In this sense, these Faraday and perpendicular configurations are complementary.  

The Voigt configuration is yet another high-symmetry configuration for which the magnetoplasmon frequency depends on the  magnetic field but again does not depend on the wave vector. In the Voigt configuration, the surface wave propagates perpendicularly to the (in-plane) external magnetic field that is parallel to the interface. Then the plasmon frequency is given by $\omega_{\mathrm{V}\pm}=\left( \sqrt{\omega_{\mathrm{c}}^{2}+2\omega_{\mathrm{p}}^{2}}\pm\omega_{\mathrm{c}}\right) /2$. Here the $\pm$ solutions correspond to the Cartesian coordinate system shown in Fig.~\ref{fig.1}:  for a forward-going SMP, we have the $+$ solution and the SMP frequency is blueshifted, while for a backward-going wave the frequency of SMP is redshifted. Clearly, in the limit of vanishing magnetic fields, the plasmon frequencies reduce to $\omega_{\mathrm{p}}$ in both configurations, as expected.

Recently, Silveirinha~\textit{et al.}~\cite{M.G.S022509} and Gangaraj~\textit{et al.}~\cite{ S.A.H.G201108} studied the propagation of SMPs on a semi-infinite magnetized electron gas  when the direction of propagation is oblique to the static magnetic field. In the electrostatic approximation they found that the frequency of the SMPs does not depend on the magnitude of the wavevector, as for the Voigt and Faraday configurations. But now in the oblique configuration the SMP frequency does depend on the angle of the wavevector with respect to the magnetic field direction~\cite{M.G.S022509}. Also, more recently, in the presence of a weakly static magnetic field gradient, we found dispersive backward and forward electrostatic waves on a cold magnetized electron gas half-space in the Faraday configuration \cite{A.M054501}.

Here we study the surface magnetoplasmons on a semi-infinite magnetized electron gas and in this oblique configuration (i.e., a mixed Faraday-Voigt configuration) in more detail. 
As an interesting result, we show that new electrostatic waves exist in such a structure due to differences between the symmetry of the media in contact, just like the Dyakonov surface waves~\cite{M.I.D714, O.T126, O.T463001}. Note that these surface waves do not exist in a semi-infinite gas plasma or a semi-infinite electron plasma in a metal without a magnetic field, and may have application in signal processing with electrostatic or slow electric waves.

\section{Basic equations for the surface magnetoplasmons}
Here we derive conditions that surface magnetoplasmons in the metal-air interface should satisfy, while in the next section we study the properties of these SMPs.  Consider a semi-infinite magnetized electron gas occupying the half-space $z<0$ in Cartesian coordinates, as shown in Fig.~\ref{fig.1}. The plane $z=0$ is the electron gas-insulator interface. Without loss of generality, we assume that the external magnetic-field vector $\textbf{B}_{0}=\pm B_{0}\textbf{e}_{x}$ 
points parallel to the $x$-axis (plus and minus signs refer to $\textbf{B}_{0}$ in the positive- and negative-$x$ directions). We will investigate the propagation of a slow electric surface wave ($\textbf{E}\approx-\nabla\Phi$) whose wavevector $\textbf{k}$
points along the interface at an angle $\theta = \arctan(k_y/k_x)$ (or $180^{\circ}-\theta$) with the magnetic field $\textbf{B}_{0}=\pm B_{0}\textbf{e}_{x}$.  
 
The electric potential may be represented in the form  \begin{figure}[tb]
  \centering
 \includegraphics[width=8.5cm,clip]{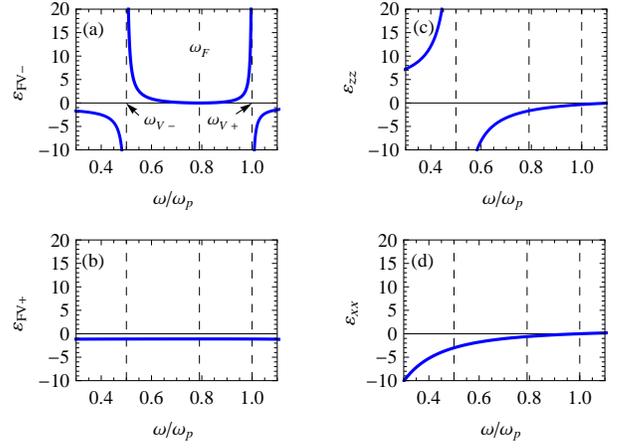}
  \caption{Variation of (a) $\varepsilon_{\mathrm{FV}-}$, (b) $\varepsilon_{\mathrm{FV}+}$, (c) $\varepsilon_{zz}$, and (d) $\varepsilon_{xx}$, with respect to the dimensionless
frequency $\omega/\omega_{\mathrm{p}}$, when $\varepsilon_{1}=1$, and $\omega_{\mathrm{c}}=0.5\omega_{\mathrm{p}}$. }
  \label{fig.2}
\end{figure}
\b 
\label{1}\Phi(x,y,z,t)=\tilde{\Phi}(z)\exp\left[ i\left( k_{x}x+ k_{y}y-\omega t\right)\right]  \;,
\e  
where $\omega$ is the angular frequency of the wave and $k_{x}$ and $k_{y}$ are wavenumbers in the $x$- and $y$-directions, respectively. For the present geometry and the magnetic field $\textbf{B}_{0}= + B_{0}\textbf{e}_{x}$, the relative dielectric tensor of the system has the form 
\b 
\label{2} \underline{\varepsilon}(\omega)=\left(%
\begin{array}{ccc}
 \varepsilon_{xx} & 0 & 0\\
  0 & \varepsilon_{yy} & \varepsilon_{yz}\\
  0 & \varepsilon_{zy} & \varepsilon_{zz}\\
\end{array}%
\right)\;,
\e
with tensor elements given by
\begin{eqnarray}
\varepsilon_{xx} & = &  1-\dfrac{\omega_{\mathrm{p}}^{2}}{\omega\left(\omega+i\gamma \right) }\;, \nonumber \\ 
\varepsilon_{yy} =  \varepsilon_{zz}  & = & 1-\dfrac{\omega_{\mathrm{p}}^{2}\left(\omega+i\gamma \right)}{\omega\left[ \left(\omega+i\gamma \right)^{2}-\omega_{\mathrm{c}}^{2}\right] }\;,\nonumber \\ 
\varepsilon_{yz} =  -\varepsilon_{zy}& = &\dfrac{i\omega_{\mathrm{c}}\omega_{\mathrm{p}}^{2}}{\omega\left[\left(\omega+i\gamma \right)^{2}-\omega_{\mathrm{c}}^{2}\right]}\;,\nonumber
\end{eqnarray}
where $\gamma$ is the damping constant. As mentioned in~\cite{M.G.S022509, S.A.H.G201108}, narrow-gap semiconductors such as InSb (with $\omega_{\mathrm{p}}/2\pi\approx4.9$THz, $\gamma/2\pi\approx0.5$THz and 
$0.25\omega_{\mathrm{p}}\leq\omega_{\mathrm{c}}\leq\omega_{\mathrm{p}}$ for a bias magnetic field in the range of  $1$ to $4$~Tesla have an optical response analogous to Eq.~\eqref{2}, where for simplicity the contribution of bound electrons to the permittivity response of InSb is disregarded, and its static (high-frequency) permittivity is taken identical to unity. Also, for our purposes, the limit of zero damping, i.e., $\gamma \to 0$ is sufficient.
Small-size limits where nonlocal response (neglected here) would start to play a role in semiconductor plasmonics including in InSb are discussed in Refs.~\cite{Maack:2017a, Maack:2018a}.

After substitution Eq.~\eqref{1} into $\nabla\cdot\textbf{D}=0$, where $\textbf{D}=\varepsilon_{0}\underline{\varepsilon}\cdot\textbf{E}$ (where $\varepsilon_{0}$ is the electric permittivity of free space), we find 
\b \label{3}
\left(\dfrac{d^{2}}{d z^{2}}-\kappa^{2}\right)\tilde{\Phi}(z)=0\;,\qquad\mbox{for} \quad z\leq0, \;\e
where  \b \label{4}\kappa=\sqrt{\dfrac{\varepsilon_{xx}k_{x}^{2}+\varepsilon_{yy}k_{y}^{2}}{\varepsilon_{zz}}}\;.
\e 
The analogous equation for the wave in the insulator is
\b \label{5}
\left(\dfrac{d^{2}}{d z^{2}}-k^{2}\right)\tilde{\Phi}(z)=0\;, \qquad \mbox{for}\quad z\geq0,
\e
with $k=\sqrt{k_{x}^{2}+k_{y}^{2}}$. Therefore, the combined solution of Eqs.~\eqref{3} and~\eqref{5} has the familiar form for a surface wave  
\b \label{6}
\tilde{\Phi}(z)=\Phi_{0}\left\{\begin{array}{clcr}
 \exp\left( -k z\right) \;, &
\mbox{$z\geq0$\;,}\
 \\ \exp\left( +\kappa z\right) \;, &
\mbox{$z\leq0$\;,}\  \\\end{array}\right.\e
where $\Phi_{0}$ is the wave amplitude.  
For the present system, the appropriate boundary condition at the separation  surface $z=0$ is
\b \label{7}
\varepsilon_{1}\dfrac{\partial\Phi_{1}}{\partial z}\bigg\vert_{z=0}=\pm\varepsilon_{zy}\dfrac{\partial\Phi_{2}}{\partial y}\bigg\vert_{z=0}+\varepsilon_{zz}\dfrac{\partial\Phi_{2}}{\partial z}\bigg\vert_{z=0}\;,
\e
where subscripts~1 and 2 refer to outside and inside the electron gas, respectively, $\varepsilon_{1}$ is the relative dielectric constant of the insulator medium and plus and minus signs refer to $\textbf{B}_{0}$ in the positive- and negative-$x$ directions, respectively.

\section{Dispersion relation and group velocities }
On applying the mentioned boundary condition~(\ref{7}) at $z=0$, we find 
\b \label{8}\varepsilon_{1}k+\varepsilon_{zz}\kappa\pm i\varepsilon_{zy}k_{y}=0\;,\e 
which leads to a relation between the wave-vector components $k_{x}$ and $k_{y}$ and the frequency $\omega$,
\b \label{9}
\dfrac{k_{y}^{2}}{k_{x}^{2}}=\varepsilon_{\mathrm{FV}\pm}(\omega)\;,
\e 
with the frequency-dependent function $\varepsilon_{\mathrm{FV}\pm}(\omega)$ to be specified shortly. The dispersion relation Eqs.~\eqref{9} is remarkable in that it does not depend on the magnitude $k$ of the wavevector, but only on the fraction $k_y/k_x$ of the wave-vector components $k_{x,y}$. In other words, the problem has a cylindrical symmetry and in agreement with Ref.~\cite{M.G.S022509} we will find that  the plasmon energies will only depend on the angle $\theta$. Yet we are interested in propagation along and perpendicular to the magnetic field and therefore will stick to Cartesian coordinates in most of what follows.

The frequency-dependent function $\varepsilon_{\mathrm{FV}\pm}(\omega)$ of Eq.~\eqref{9} has the form
\begin{figure}[tb]
  \centering
\includegraphics[width=8cm,clip]{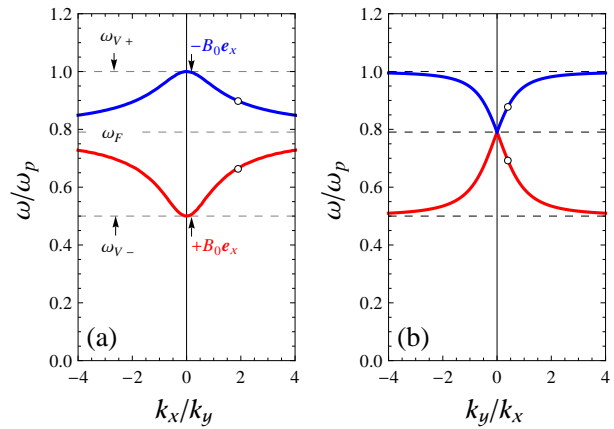}
  \caption{Dispersion curves of SMPs at a flat magnetized electron gas-vacuum interface, as obtained from Eq.~\eqref{9} for $\omega_{\mathrm{c}}=0.5\omega_{\mathrm{p}}$. Panel~(a): $ k_{y}$ is constant (positive). Panel~(b): $k_{x}$ is constant (positive). }
  \label{fig.3}
\end{figure} 

\begin{eqnarray}
\varepsilon_{\mathrm{FV}\pm} &=& \dfrac{-B\pm\sqrt{B^{2}-4AC}}{2A}\;, \qquad \mbox{with} \nonumber \\
A &=& \left(\varepsilon_{1}^{2}-\varepsilon_{zy}^{2}-\varepsilon_{yy}\varepsilon_{zz} \right)^{2}+4\varepsilon_{1}^{2}\varepsilon_{zy}^{2}\;, \nonumber \\ 
B &=& 2\left(\varepsilon_{1}^{2}-\varepsilon_{zy}^{2}-\varepsilon_{yy}\varepsilon_{zz} \right)\left(\varepsilon_{1}^{2}-\varepsilon_{xx}\varepsilon_{zz} \right)+4\varepsilon_{1}^{2}\varepsilon_{zy}^{2}\;, \nonumber \\
C &=&\left(\varepsilon_{1}^{2}-\varepsilon_{xx}\varepsilon_{zz} \right)^{2}\;. \nonumber
\end{eqnarray}
It is clear that the frequency dependence of $\varepsilon_{\mathrm{FV}\pm}(\omega)$ originates fully from the assumed frequency dispersion of the dielectric functions. 

We will now look for propagating-wave solutions in the $x$- and $y$-directions, so that
the right-hand side of Eq.~\eqref{9} must be positive-valued. Working with the reduced variables $\omega/\omega_{\mathrm{p}}$ and $\omega_{\mathrm{c}}/\omega_{\mathrm{p}}$, we depict in Fig.~\ref{fig.2}  the variation of (a) $\varepsilon_{\mathrm{FV}-}$, (b) $\varepsilon_{\mathrm{FV}+}$, (c) $\varepsilon_{zz}$, and (d) $\varepsilon_{xx}$, with respect to the
dimensionless frequency $\omega/\omega_{\mathrm{p}}$, when $\varepsilon_{1}=1$ and $\omega_{\mathrm{c}}=0.5\omega_{\mathrm{p}}$. From Fig.~\ref{fig.2}(a,b) it then follows that only the lower-branch solution $\varepsilon_{\mathrm{FV} -}(\omega)$  
can lead to propagating-wave solutions in Eq.~\eqref{9}, and only in a finite range of frequencies. Furthermore, for having a surface wave, $\varepsilon_{zz}$ in panel~(c) must be negative. There is indeed a region below and above the SMP frequencies in the Voigt configuration, i.e., $\omega_{\mathrm{V}-}=0.5\omega_{\mathrm{p}}$ and $\omega_{\mathrm{V}+}=\omega_{\mathrm{p}}$, where $\varepsilon_{\mathrm{FV}-}$ is positive and $\varepsilon_{zz}$ is negative, and it is there that the conditions for propagating SMPs are satisfied. In this region, $\varepsilon_{xx}$ is also negative, as shown in panel~(d). Note that the existence of these SMPs is due to differences between the anisotropy of the two media, similar to Dyakonov surface waves~\cite{M.I.D714, O.T126, O.T463001}. However, Dyakonov surface waves are a type of electromagnetic waves localized at an interface between two transparent media: an isotropic medium and a uniaxial crystal. In contrast to the Dyakonov surface waves, here SMPs are a type of electrostatic (or, more accurately, quasi-electrostatic) waves localized at an interface between an isotropic medium and an electric-gyrotropic electron plasma~\cite{A.M2947}.

In Fig.~\ref{fig.3}(a) we show the variation of $\omega/\omega_{\mathrm{p}}$ with respect to $k_{x}/k_{y}$, for a positive constant value of $k_{y}$, while in panel~(b) we show the same as a function of  $k_{y}/k_{x}$, for a positive constant value of $k_{x}$. Two curves appear in each panel in agreement with $\textbf{B}_{0}=\pm B_{0}\textbf{e}_{x}$. Note that for general dispersion relations it would be important to state the value of the wavevector that is kept constant, but not so for our dispersion relation where the dependence is only on the fraction $k_{y}/k_{x}$ (or $k_{x}/k_{y}$). The panels in Fig.~\ref{fig.3} give complementary information. Alternatively, one could plot the frequencies as a function of the angle $\theta$ (not shown).  

If we consider $k_{y}$ to be fixed and positive and take the symbol \textquotedblleft $+$\textquotedblright\,  in Eq.~\eqref{8} (i.e., $\textbf{B}_{0}=+B_{0}\textbf{e}_{x}$), then for $k_{x}>0$ ($k_{x}<0$) there is a SMP with $v_{\mathrm{g}x}>0$ ($v_{\mathrm{g}x}<0$) in the region below the line $\omega=\omega_{\mathrm{F}}=\sqrt{\omega_{\mathrm{p}}^{2}+\omega_{\mathrm{c}}^{2}}/\sqrt{2}$ and above the line $\omega=\omega_{\mathrm{V}-}$ (see red curve in panel~(a) of Fig.~\ref{fig.3}). For $v_{\mathrm{g}x}>0$, we note that $k_{x}$ and $k_{y}$ are both positive, while for $v_{\mathrm{g}x}<0$, we have $\left(-k_{x},+k_{y} \right)$. 

If we consider $k_{y}$ to be fixed and positive and take the symbol \textquotedblleft $-$\textquotedblright \;  in Eq.~\eqref{8} (i.e., $\textbf{B}_{0}=-B_{0}\textbf{e}_{x}$), then for $k_{x}>0$ ($k_{x}<0$) there is a SMP with $v_{\mathrm{g}x}<0$ ($v_{\mathrm{g}x}>0$) in the region below the line $\omega=\omega_{\mathrm{V}+}$ and above the line $\omega=\omega_{\mathrm{F}}$ (blue curve in panel~(a) of Fig.~\ref{fig.3}). Again, for $v_{\mathrm{g}x}<0$ one can find that $k_{x},k_{y}>0$, while for $v_{\mathrm{g}x}>0$, we obtain $\left(-k_{x},+k_{y} \right)$.

If we consider $k_{x}$ to be fixed with positive sign and take the symbol \textquotedblleft $+$\textquotedblright \;  in Eq.~\eqref{8} (i.e., $\textbf{B}_{0}=+B_{0}\textbf{e}_{x}$), then there is a SMP with $v_{\mathrm{g}y}<0$ ($v_{\mathrm{g}y}>0$ ) in the region below the line $\omega=\omega_{\mathrm{F}}$ and above the line $\omega=\omega_{\mathrm{V}-}$ (see red curve in panel~(b) of Fig.~\ref{fig.3}) for $k_{y}>0$ ($k_{y}<0$). For $v_{\mathrm{g}y}<0$, we have $\left(+k_{x},+k_{y} \right)$, while $\left(+k_{x}, -k_{y} \right)$  yields $v_{\mathrm{g}y}>0$. 

Finally, if we consider $k_{x}$ to be fixed with positive sign and take the symbol \textquotedblleft $-$\textquotedblright \;  in Eq.~\eqref{8} (i.e., $\textbf{B}_{0}=-B_{0}\textbf{e}_{x}$), then there is a SMP with $v_{\mathrm{g}y}>0$ ($v_{\mathrm{g}y}<0$) in the region below the line $\omega=\omega_{\mathrm{V}+}$ and above the line $\omega=\omega_{\mathrm{F}}$ (blue curve in panel~(b) of Fig.~\ref{fig.3}). For $\left(+k_{x},+k_{y} \right)$ we see $v_{\mathrm{g}y}>0$, while it is clear that $v_{\mathrm{g}y}$ is negative for $\left(+k_{x},-k_{y} \right)$ .
\begin{figure}[tb]
  \centering
\includegraphics[width=8cm,clip]{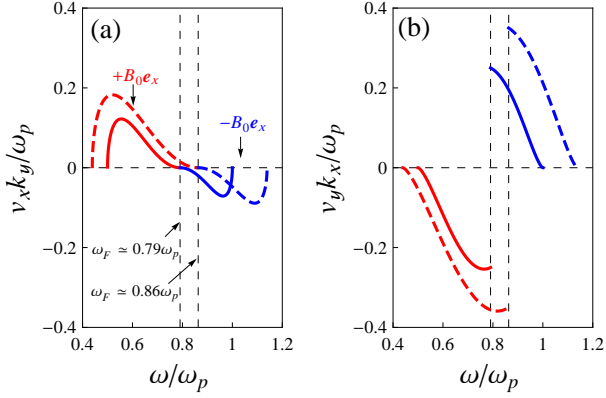}
  \caption{Group (energy) velocity curves of SMPs at a flat magnetized electron gas-vacuum interface, as obtained from Eqs.~\eqref{10} and~\eqref{11}, when $k_{x}$ and $k_{y}$ are positive. Panel~(a): variation of $v_{x}$ (i.e., $v_{\mathrm{g}x}$ or $v_{\mathrm{e}x}$), when $k_{y}$ is constant. Panel~(b): variation of $v_{y}$ (i.e., $v_{\mathrm{g}y}$ or $v_{\mathrm{e}y}$), when $k_{x}$ is constant. Here the different curves refer to  $\omega_{\mathrm{c}}=0.5\omega_{\mathrm{p}}$ (solid lines), and $\omega_{\mathrm{c}}=0.7\omega_{\mathrm{p}}$ (dashed lines).}
  \label{fig.4}
\end{figure} 

To examine the conditions for the validity of these modes we calculate the components of the group velocity of the new waves by using Eq.~\eqref{8}. We 
differentiate the equation as it stands, first with respect to $k_{x}$ while keeping $k_{y}$ constant, and then with respect to $k_{y}$ keeping $k_{x}$ constant. In doing so, we have to remember that $\omega$ is a function of both $k_{x}$ and $k_{y}$. We find the group-velocity components
%
\b 
\label{10} v_{\mathrm{g}x}=\dfrac{k_{x}\left(\dfrac{\varepsilon_{1}}{k}+\dfrac{\varepsilon_{xx}}{\kappa} \right) }{\mp i k_{y}\dfrac{d \varepsilon_{zy}}{d \omega}-\kappa\dfrac{d \varepsilon_{zz}}{d \omega}-\dfrac{k_{x}^{2}}{2\kappa}\left(\dfrac{d\varepsilon_{xx}}{d\omega}-\dfrac{\varepsilon_{xx}}{\varepsilon_{zz}}\dfrac{d\varepsilon_{zz}}{d\omega} \right) }\;, \e
\b \label{11} v_{\mathrm{g}y}=\dfrac{\mp i\varepsilon_{zy}-k_{y}\left(\dfrac{\varepsilon_{1}}{k}+\dfrac{\varepsilon_{yy}}{\kappa} \right) }{\pm i k_{y}\dfrac{d \varepsilon_{zy}}{d \omega}+\dfrac{\kappa}{2}\dfrac{d \varepsilon_{zz}}{d \omega}+\dfrac{1}{2\kappa}\left(k_{x}^{2}\dfrac{d\varepsilon_{xx}}{d\omega}+k_{y}^{2}\dfrac{d\varepsilon_{yy}}{d\omega} \right) }\;. 
\e 
To better understand the behavior of the SMPs, we show the variation of these group-velocity components with respect to $\omega/\omega_{\mathrm{p}}$ in Fig.~\ref{fig.4}, using the same parameter values as for Fig.~\ref{fig.2} (solid lines), when $k_{x}$ and $k_{y}$ are positive. The dashed lines show the result for the case $\omega_{\mathrm{c}}=0.7\omega_{\mathrm{p}}$.  Panel~\ref{fig.4}(a) shows the behavior of the dimensionless variable $v_{\mathrm{g}x}k_{y}/\omega_{\mathrm{p}}$ and panel~(b)  the variation of $v_{\mathrm{g}y}k_{x}/\omega_{\mathrm{p}}$. Both group-velocity components change sign at $\omega = \omega_{\mathrm{F}}$, but in a very different fashion: $v_{\mathrm{g}x}$ goes through zero continuously, whereas $v_{\mathrm{g}y}$ makes a discontinous jump. 

\begin{figure}[tb]
  \centering
\includegraphics[width=8cm,clip]{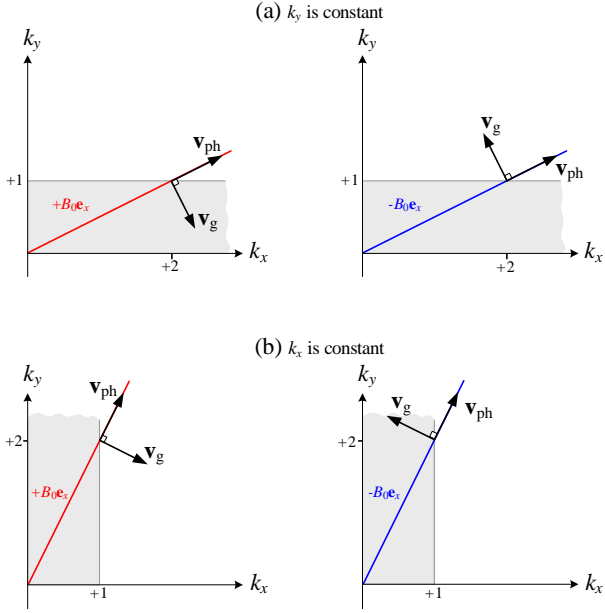}
  \caption{Iso-frequency curves $(\omega=\mathrm{constant})$ of SMPs at a flat magnetized electron gas-vacuum interface, as obtained from Eq.~\eqref{9} corresponding to the labeled points in Fig.~\ref{fig.3}, for $+B_{0}\textbf{e}_{x}$ (left figures) and $-B_{0}\textbf{e}_{x}$ (right figures). Panel~(a): $k_{y}$ is constant. Panel~(b): $k_{x}$ is constant. It is clear that the vectors $\textbf{v}_{\mathrm{ph}}$ and $\textbf{v}_{\mathrm{g}}$ are always perpendicular, and both are in the $xy$-plane. So the outer product of these two vectors always points in the $\pm z$-direction. Note that for $+B_{0}\textbf{e}_{x}$ ($-B_{0}\textbf{e}_{x}$), the outer product $\textbf{v}_{\mathrm{ph}}\times\textbf{v}_{\mathrm{g}}$ points in the electron plasma (vacuum) direction.}
  \label{fig.5}
\end{figure}

In an anisotropic medium the direction of group (signal) propagation differs in general from the direction of phase propagation. Indeed, the phase velocity  points along the wave vector $\textbf{k}=k_{x}\textbf{e}_{x}+k_{y}\textbf{e}_{y}$, whereas the energy propagates in the direction of the Poynting vector and the signal velocity. The magnitude of the phase velocity is given by $v_{\mathrm{ph}}=\omega/k$, and the phase-velocity vector makes an angle $\theta_{\mathrm{ph}}=\theta=\arctan\left(k_{y}/k_{x} \right) $ with $\textbf{B}_{0}=+ B_{0}\textbf{e}_{x}$ (or the $x$-axis). But what is the direction of the group velocity vector $\nabla_{\bf k} \omega$ ? If we define $\theta_{\mathrm{g}}$ to be the angle between the group-velocity and the magnetic-field directions, then we can obtain $\tan(\theta_{\mathrm{g}})$ as $v_{\mathrm{g}y}/v_{\mathrm{g}x}$, i.e.  by dividing Eq.~\eqref{11} by Eq.~\eqref{10}. After using Eqs.~\eqref{4} and~\eqref{8}, we find that \b \label{12} \tan~\theta_{\mathrm{g}}=-\dfrac{k_{x}}{k_{y}}=-\dfrac{1}{\tan~\theta_{\mathrm{ph}}}.  
\e The phase and group velocities of the quasi-electrostatic magnetoplasmon waves  therefore have the remarkable property that they are perpendicular to each
other: {\em energy flows in the direction perpendicular to the phase propagation}. This property is shown in Fig.~\ref{fig.5} and reflects the dispersion relation~\eqref{8} for which $\omega(k,\theta) = \omega(\theta)$: the group velocity in the $\bf k$-direction vanishes identically irrespective of the angle between $\bf k$ and the magnetic field, while the group velocity  perpendicular to $\bf k$  is finite. In optics the situation of group velocities being exactly opposite to their phase velocities is well known to  occur for negative-index materials~\cite{Veselago:1968a, Pendry:2000a, Smith:2004a}. Perpendicular group and phase velocities on the other hand for all wave vectors as found here are less well known in optics. However, in fluid dynamics there is an interesting analogy with so-called \textquotedblleft internal waves'' (or \textquotedblleft(internal) gravity waves''). These are well known to have perpendicular group and phase velocities whatever their angle with the water surface, see for example Ref.~\cite{Kundu:2016a}.
\begin{figure}[!htb]
  \centering
 \includegraphics[width=8cm,clip]{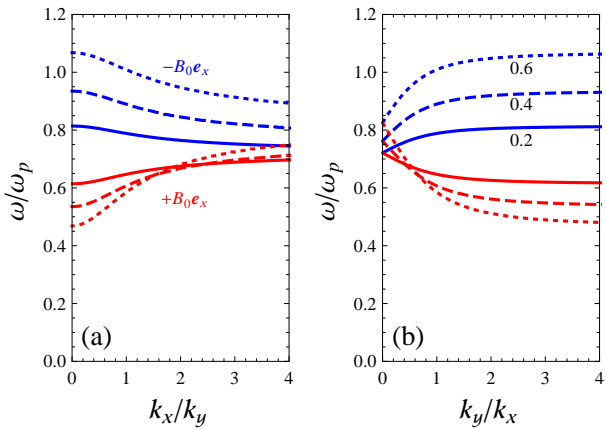}
  \caption{Dispersion curve of SMPs at a flat magnetized electron gas-vacuum interface, as obtained from Eq.~\eqref{9}, when (a) $k_{y}$ is positive constant and (b) $k_{x}$ is positive constant. Here the different curves correspond to  $\omega_{\mathrm{c}}=0.2\omega_{\mathrm{p}}$ (solid lines), $\omega_{\mathrm{c}}=0.4\omega_{\mathrm{p}}$ (dashed lines), and $\omega_{\mathrm{c}}=0.6\omega_{\mathrm{p}}$ (dotted lines). }
  \label{fig.6}
\end{figure}

We now turn to the effect of the magnetic-field strength through the cyclotron frequency~$\omega_{\mathrm{c}}$. Again, Fig.~\ref{fig.6} is a plot of $\omega/\omega_{\mathrm{p}}$ versus (a) $k_{x}/k_{y}$ and (b) $k_{y}/k_{x}$. Here the different curves refer to  $\omega_{\mathrm{c}}=0.2\omega_{\mathrm{p}}$ (solid lines), $\omega_{\mathrm{c}}=0.4\omega_{\mathrm{p}}$ (dashed lines), and $\omega_{\mathrm{c}}=0.6\omega_{\mathrm{p}}$ (dotted lines). One can see that changing the parameter $\omega_{\mathrm{c}}$ has a strong effect on the dispersion curve of the SMPs. For weaker magnetic fields, the dispersion curves are flatter and group velocities smaller. Thus group velocities of the SMPs can be controlled with the static magnetic field as the control parameter, see Fig.~\ref{fig.4}.  From panel~(a), it is clear that for the $x$-forward mode (when $k_{x},k_{y} >0$), increasing  $\omega_{\mathrm{c}}$ redshifts the frequency of the mode for low values of $k_{x}/k_{y}$ and blueshifts the mode frequency for high values of $k_{x}/k_{y}$. From the same  panel~(a), it follows that for the $x$-backward mode and $k_{x},k_{y} >0$, increasing $\omega_{\mathrm{c}}$ blueshifts the SMP frequency. Finally, as mentioned before, one constant-frequency solution $\omega_{\mathrm{F}}$ exists in the Faraday geometry, while two solutions $\omega_{\mathrm{V}\pm}$ exist in the Voigt geometry. As can be seen in panel~\ref{fig.6}(a), in the limit 
$k_{x}/k_{y}\rightarrow0$, we indeed obtain the two frequencies of the Voigt geometry, i.e., $\omega=\omega_{\mathrm{V}-}$, when we take $+B_{0}\textbf{e}_{x}$; and $\omega=\omega_{\mathrm{V}+}$, when we consider the $-B_{0}\textbf{e}_{x}$ solution. Also, from the limit $k_{y}/k_{x}\rightarrow0$ in Fig.~\ref{fig.6}(b), we indeed find back the result for the Faraday geometry, i.e., $\omega=\omega_{\mathrm{F}}$. 

\section{Power flow of a surface magnetoplasmon}
\label{IV}
In this section, we calculate the $x$- and $y$-components of the power flow of the SMPs, i.e. along and perpendicular to the magnetic field, respectively. Under the electrostatic approximation~\cite{Maier:2007a, A.M2947, A.M2976}, the power flow associated with the SMPs of a semi-infinite magnetized electron gas is given by
\begin{figure}[tb]
  \centering
 \includegraphics[width=8cm,clip]{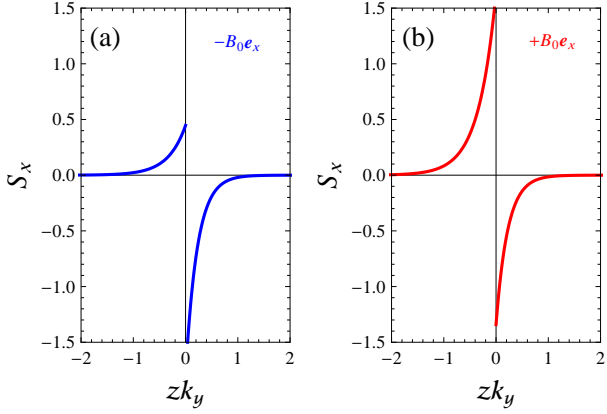}
  \caption{Normalized profile $S_{x}(z)$ of SMP modes of a flat magnetized electron gas-vacuum interface, as obtained from Eq.~\eqref{16} when $k_{x}/k_{y}=2$, $k_{x}>0$, and $ k_{y}$ is a positive constant, corresponding to the labeled points in Fig.~\ref{fig.3}(a). Panel~(a): $\textbf{B}_{0}=-B_{0}\textbf{e}_{x}$. Panel~(b):  $\textbf{B}_{0}=+B_{0}\textbf{e}_{x}$.   }
  \label{fig.7}
\end{figure}
\b \label{13}
\textbf{S}=-\varepsilon_{0}\left\{\begin{array}{clcr}
\varepsilon_{1}\Phi_{1}\dfrac{\partial}{\partial t}\nabla\Phi_{1}\;, &
\mbox{$z>0$\;,}\
\\  \Phi_{2}\dfrac{\partial}{\partial t}\left[ \underline{\varepsilon}(\omega)\cdot\nabla\Phi_{2}\right] \;, &
\mbox{$z< 0$\;,}\ 
  \\ \end{array}\right.
\e
where these vectors in the two media have components in the $x$-, $y$- and $z$-directions. The cycle-averaged $x$- and $y$-components are
\begin{figure}[tb]
  \centering
 \includegraphics[width=8cm,clip]{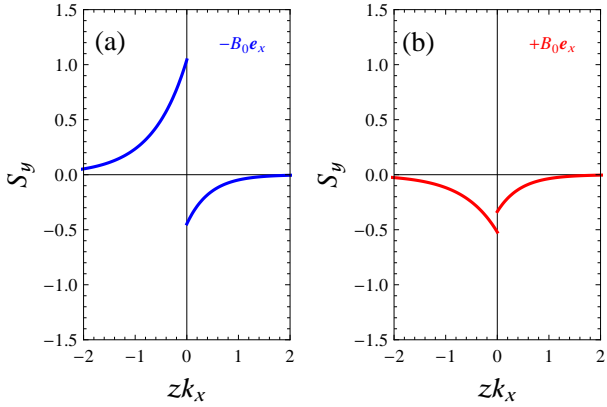}
  \caption{Normalized profile $S_{y}(z)$ of SMP modes of a flat magnetized electron gas-vacuum interface, as obtained from Eq.~\eqref{17} when $k_{y}/k_{x}=0.5$, $k_{y}>0$, and $ k_{x}$ has a positive constant value, corresponding to the labeled points in Fig.~\ref{fig.3}(b). Panel (a): $\textbf{B}_{0}=-B_{0}\textbf{e}_{x}$. Panel (b):  $\textbf{B}_{0}=+B_{0}\textbf{e}_{x}$. }
  \label{fig.8}
\end{figure}  \b \label{14}
S_{x}=-\dfrac{\varepsilon_{0}}{2}{\rm Re}\left\{\begin{array}{clcr}
\varepsilon_{1}\Phi_{1}\dfrac{\partial}{\partial t}\dfrac{\partial}{\partial x}\Phi_{1}^{\ast}\;, &
\mbox{$z>0$\;,}\
 \\ \varepsilon_{xx}\Phi_{2}\dfrac{\partial}{\partial t}\dfrac{\partial}{\partial x}\Phi_{2}^{\ast}\;, &
\mbox{$ z< 0$\;,}\ 
  \\ \end{array}\right.\e
\begin{equation} \label{15}
S_{y}=-\dfrac{\varepsilon_{0}}{2}{\rm Re}\left\{\begin{array}{clcr}
\varepsilon_{1}\Phi_{1}\dfrac{\partial}{\partial t}\dfrac{\partial}{\partial y}\Phi_{1}^{\ast}\;, &
\mbox{$z>0$\;,}\
 \\  \Phi_{2}\dfrac{\partial}{\partial t}\left( \varepsilon_{yy}\dfrac{\partial}{\partial y}\mp\varepsilon_{yz}\dfrac{\partial}{\partial z}\right) \Phi_{2}^{\ast}\;, &
\mbox{$ z< 0$\;,}\ 
  \\ \end{array}\right. \end{equation}
in the complex-number representation, where $^{\ast}$ denotes complex conjugation, and ${\rm Re}$ denotes taking the real part. Note that $\varepsilon_{xx}$ and $\varepsilon_{yy}$ are real. After substitution of Eq.~\eqref{6} into Eqs.~\eqref{14} and~\eqref{15} and using Eq.~\eqref{1}, we obtain
\b \label{16}
S_{x}=-\dfrac{1}{2}\varepsilon_{0}\omega k_{x}\Phi_{0}^{2}\left\{\begin{array}{clcr}
\varepsilon_{1} e^{-2k z}\;, &
\mbox{$z>0$\;,}\ 
 \\ \varepsilon_{xx} e^{+2\kappa z}\;, &
\mbox{$z<0$\;,}\ \\ \end{array}\right.\ \e
\b \label{17}
S_{y}=-\dfrac{1}{2}\varepsilon_{0}\omega k_{y}\Phi_{0}^{2}\left\{\begin{array}{clcr}
\varepsilon_{1} e^{-2k z}\;, &
\mbox{$z>0$\;,}\ 
 \\ \left( \varepsilon_{yy}\mp i\varepsilon_{yz}\dfrac{\kappa}{k_{y}}\right)  e^{+2\kappa z}\;, &
\mbox{$z< 0$\;,}\  \\ \end{array}\right.\ \e
We note that the distributions in Eqs.~\eqref{16} and~\eqref{17} are discontinuous at the interface $z=0$. The total power flow densities (per unit width), associated with the SMPs can be determined by an integration over $z=0$. We find 
\b \label{170}
\left\langle S_{x}\right\rangle =-\dfrac{1}{4}\varepsilon_{0}\omega k_{x}\left[\dfrac{\varepsilon_{1}}{k}+\dfrac{\varepsilon_{xx}}{\kappa} \right] \Phi_{0}^{2} \;,\e
\b \label{180}
\left\langle S_{y}\right\rangle =-\dfrac{1}{4}\varepsilon_{0}\omega k_{y}\left[\dfrac{\varepsilon_{1}}{k}+\dfrac{1}{\kappa}\left( \varepsilon_{yy}\mp i\varepsilon_{yz}\dfrac{\kappa}{k_{y}}\right)  \right] \Phi_{0}^{2} \;,\e
where $\left\langle \cdots\right\rangle\equiv\int_{-\infty}^{+\infty}\cdots dz$. In Fig.~\ref{fig.7}, we calculate the normalized profiles of $S_{x}(z)$ of SMP modes of a flat magnetized electron gas-vacuum interface,
when $k_{x}/k_{y}=2$, $k_{x}>0$, and $ k_{y}$ is positive constant, corresponding to the labeled points in Fig.~\ref{fig.3}(a). It can be seen
that power flow densities are largest at the boundary, and their amplitudes decay
exponentially with increasing distance into each medium from the interface. Comparing the curve of case $\textbf{B}_{0}=-B_{0}\textbf{e}_{x}$ in Fig.~\ref{fig.3}(a) by the result in Fig.~\ref{fig.7}(a), we conclude that the $x$-backward SMP mode in the region below the line $\omega=\omega_{\mathrm{V}+}$ and above the line $\omega=\omega_{\mathrm{F}}$ in panel~(a) of Fig.~\ref{fig.3} (blue curve) is an acceptable mode with electric potential shown by Eq. \eqref{1}. For the case $\textbf{B}_{0}=+B_{0}\textbf{e}_{x}$ in the region below the line $\omega=\omega_{\mathrm{F}}$ and above the line $\omega=\omega_{\mathrm{V}-}$ of panel~(a) of Fig.~\ref{fig.3} (red curve), the power flow in the EG region occurs in the $+x$-direction, while in the vacuum region, the power flow occurs in the $-x$-direction, i.e., opposite to the direction of phase propagation. Also, the total power flow density (per unit width) is positive for the $x$-forward SMP mode. This result is in agreement with the behavior of dispersion curve of the $x$-forward SMP, shown in Fig.~\ref{fig.3}(a) and thus we have again an acceptable mode. 
  
In Fig.~\ref{fig.8}, by using Eq.~\eqref{17}, we calculate the
normalized profiles of $S_{y}(z)$ of SMP modes of a flat magnetized electron gas-vacuum interface,
when $k_{y}/k_{x}=0.5$, $k_{y}>0$, and $ k_{x}$ is positive constant, corresponding to the labeled points in Fig.~\ref{fig.3}(b). Here, one can see that for the $y$-forward SMP mode (panel~(a) of Fig.~\ref{fig.8}), the power flow in the EG region occurs in the $+y$-direction, while in the vacuum region, the power flow occurs in the $-y$-direction. Also, we find that the total power flow density (per unit width) is positive for the $y$-forward mode. This result is in agreement with the behavior of dispersion curve of $y$-forward SMP modes in Fig.~\ref{fig.3}(b) for the case $\textbf{B}_{0}=-B_{0}\textbf{e}_{x}$. Furthermore, comparing the lower curve in Fig.~\ref{fig.3}(b) for the case $\textbf{B}_{0}=+B_{0}\textbf{e}_{x}$ by the result in Fig.~\ref{fig.8}(b), we conclude that the $y$-backward SMP mode in the region below the line $\omega=\omega_{\mathrm{F}}$ and above the line $\omega=\omega_{\mathrm{V}-}$ in panel~(b) of Fig.~\ref{fig.3} (red curve) is also an acceptable mode with electric potential shown by Eq. \eqref{1}. Finally, it is clear from Figs. \ref{fig.7} and \ref{fig.8} that the power in the upper and lower half spaces flows in different directions, but not for $S_{y}$ in the case $\textbf{B}_{0}=+B_{0}\textbf{e}_{x}$. Actually, for the $y$-backward SMP mode the power flow in both media occurs in the $-x$-direction.

\section{Energy distribution and energy velocities}
Now, we first consider the energy distribution in the transverse direction. For the cycle-averaged energy distribution associated with the SMPs of a semi-infinite electron gas, we have, in the two media 
\b \label{20}
U=\dfrac{\varepsilon_{0}}{4}\left\{\begin{array}{clcr}
\varepsilon_{1}\vert\nabla\Phi_{1}\vert^{2}\;, &
\mbox{$z>0$\;,}\
 \\  \nabla\Phi_{2}^{\ast}\cdot\left( \dfrac{d\left( \omega\underline{\varepsilon}\right) }{d\omega} \cdot\nabla\Phi_{2}\right) \;, &
\mbox{$ z< 0$\;,}\ 
  \\ \end{array}\right.
  \e
where losses are neglected. After substitution Eq.~\eqref{6} into~\eqref{20}, we obtain 
\begin{multline} \label{21}
U=\dfrac{1}{4}\varepsilon_{0}\Phi_{0}^{2}\\ \times\left\{\begin{array}{clcr}
2\varepsilon_{1}k^{2}e^{-2kz}\;, &
\mbox{$z>0$\;,}\
 \\ \left( k_{x}^{2} \dfrac{d\left( \omega\varepsilon_{xx}\right) }{d\omega}+k_{y}^{2}\dfrac{d\left( \omega\varepsilon_{yy}\right) }{d\omega}+\Xi\right) e^{+2\kappa z}  \;, &
\mbox{$z< 0$\;,}\ 
 \\ \end{array}\right. 
 \end{multline}
in the complex-number representation, where $$\Xi=\kappa^{2}\dfrac{d\left( \omega\varepsilon_{zz}\right) }{d\omega} \mp 2i\kappa k_{y}\dfrac{d\left( \omega\varepsilon_{yz}\right) }{d\omega} \;.$$ From Eq.~\eqref{21} we find that the  contributions to the energy density of the two half spaces are both positive. The total energy density associated with the SMPs is again determined by integration over the out-of-plane coordinate $z$, the energy per unit surface area being
\begin{multline} \label{22}
\left\langle U\right\rangle =\dfrac{1}{4}k\varepsilon_{0}\\\times \left[ \varepsilon_{1}+\dfrac{ 1}{2k\kappa}\left( k_{x}^{2} \dfrac{d\left( \omega\varepsilon_{xx}\right) }{d\omega}+k_{y}^{2}\dfrac{d\left( \omega\varepsilon_{yy}\right) }{d\omega}+\Xi\right)\right] \Phi_{0}^{2} \;.\end{multline}
In general, the energy velocity of the SMPs is given as the ratio of the total power flow density
(per unit width) and the total energy density (per unit area). For our model of the SMPs, this leads to the energy-velocity components 
\b \label{23} v_{\mathrm{e}x}=-\dfrac{\omega k_{x}}{k}\dfrac{\dfrac{\varepsilon_{1}}{k}+\dfrac{\varepsilon_{xx}}{\kappa}  }{\varepsilon_{1}+\dfrac{ 1}{2k\kappa}\left( k_{x}^{2} \dfrac{d\left( \omega\varepsilon_{xx}\right) }{d\omega}+k_{y}^{2}\dfrac{d\left( \omega\varepsilon_{yy}\right) }{d\omega}+\Xi\right)}\;, \e
\b \label{24} v_{\mathrm{e}y}=-\dfrac{\omega k_{y}}{k}\dfrac{\dfrac{\varepsilon_{1}}{k}+\dfrac{1}{\kappa}\left( \varepsilon_{yy}\mp i\varepsilon_{yz}\dfrac{\kappa}{k_{y}}\right)  }{\varepsilon_{1}+\dfrac{ 1}{2k\kappa}\left( k_{x}^{2} \dfrac{d\left( \omega\varepsilon_{xx}\right) }{d\omega}+k_{y}^{2}\dfrac{d\left( \omega\varepsilon_{yy}\right) }{d\omega}+\Xi\right) }\;, \e
The expression on the right-hand sides of Eqs.~\eqref{23} and~\eqref{24} are precisely those obtained from the usual
definition of the group velocity of SMPs in the absence of damping, i.e., Eqs.~\eqref{10} and~\eqref{11}, as were shown in Fig.~\ref{fig.4}. This means that the net power flow is in the direction of the group velocity. Let us note that, by contrast, in resonant multiply scattering media, the group and transport velocities in general will differ \cite{A.L143}.

\section{Conclusions}
In summary, we have studied the propagation of SMPs on a semi-infinite magnetized
electron gas in the electrostatic approximation by consideration of a mixed Faraday-Voigt configuration. We have shown that such a structure permits propagation of new electrostatic waves that are strongly direction-dependent and do
not exist in a semi-infinite gas plasma or a semi-infinite electron plasma in a metal. We have studied the dispersion relation, group velocity and energy relations of the found magnetoplasmons in detail.
In particular, we found that the group velocities of the SMPs can be controlled by the applied static magnetic field and that the phase and group velocities are always perpendicular for these surface magnetoplasmons. Furthermore, we analyzed situations in which power will flow in different directions in the upper and lower half spaces, while we also discussed cases where  in the upper and lower half spaces the power will flow in the same directions.

\section*{Funding}
M.W. acknowledges support by the Danish National Research Foundation through NanoPhoton - Center for
Nanophotonics (grant number DNRF147) and Center
for Nanostructured Graphene (grant number DNRF103), and from the Independent Research
Fund Denmark - Natural Sciences (project no.
0135-004038).

\section*{Conflicts of interest/Competing interests}
The authors declare no conflicts of interest/competing interests.

\section*{Authors' contributions}
A. M. proposed the idea and performed the initial calculations and analyzed the initial numerical data. Both authors (A. M. and M. W.) have discussed the results thoroughly and contributed to the writing and review of the manuscript.

\section*{Additional information}
Correspondence and requests for materials should be addressed to A.M or M.W.

\end{document}